\documentclass[a4paper]{jpconf}
\usepackage{graphicx}
\usepackage{lineno}
\usepackage{citesort}
\begin{document}
\title{Measurement of D-meson production in Pb--Pb collisions at the LHC with ALICE}

\author{Andrea Festanti for the ALICE Collaboration}

\address{Universit\`a degli Studi di Padova and INFN Sezione di Padova, via Marzolo 8, 35131 Padova, Italy}

\ead{andrea.festanti@pd.infn.it}

\begin{abstract}
The measurement of D-meson production in heavy-ion collisions at LHC energy provides insights into the mechanisms of interaction of charm quarks in the hot and dense medium formed in these collisions. ALICE results on the D-meson nuclear modification factor and azimuthal anisotropy in Pb--Pb collisions at $\sqrt{s_{\rm NN}}=2.76$~TeV are presented.  
\end{abstract}

\section{Introduction}

Heavy-flavour hadrons are suitable to probe the conditions of the high-energy-density Quark-Gluon Plasma (QGP) medium formed in ultra-relativistic heavy-ion collisions. Heavy quarks are mainly produced in hard scattering processes in the early stage of Pb--Pb collisions. The time-scale of their production ($\Delta\tau< 1/2m_{c,b}\sim 0.07~{\rm fm}/c~{\rm for~charm~and}\sim 0.02~{\rm fm}/c~{\rm for~beauty}$) is, in general, shorter than the formation time of the QGP, $\tau_{0}\sim0.1-1$~fm/$c$. During their propagation through the medium, heavy quarks interact with its constituents and lose energy. QCD energy loss is expected to occur via both inelastic (radiative energy loss via medium-induced gluon radiation)~\cite{Baier1997265} and elastic (collisional energy loss) processes~\cite{PhysRevD.44.1298}. The energy loss for quarks is expected to be smaller than for gluons, due to the smaller colour coupling factor of quarks with respect to gluons. In addition, the ``dead-cone effect'' should reduce small-angle gluon radiation for heavy quarks with moderate energy compared to their mass, thus further attenuating the effect of the medium~\cite{PhysRevD.69.114003}.  

The nuclear modification factor $R_{\rm AA}(p_{\rm T})=({\rm d}N_{\rm AA}/{\rm d}p_{\rm T})/(\langle T_{\rm AA}\rangle\cdot{\rm d}\sigma_{\rm pp}/{\rm d}p_{\rm T})$, where $\langle T_{\rm AA}\rangle$ is the average nuclear overlap function for a given centrality class, is sensitive to the interaction of hard partons with the medium. At large $p_{\rm T}$, $R_{\rm AA}$ is expected to be mostly sensitive to the average energy loss of heavy-quarks in the hot and dense medium. The questions whether low-momentum heavy quarks can reach thermal equilibrium with the medium constituents and participate in the collective expansion of the system~\cite{Batsouli200326,Greco2004202}, and whether heavy quarks can hadronise also via recombination with other quarks from the medium~\cite{Greco2004202,Andronic200336} are still open. 
These questions are addressed by studying $R_{\rm AA}$ at low and intermediate $p_{\rm T}$ and measuring the azimuthal anisotropy of heavy-flavour hadron production with respect to the reaction plane, defined by the beam axis and the impact parameter of the collision. The hadronisation mechanisms of the c quark are also investigated through the measurement of  ${\rm D}_{\rm s}^+$ production in nucleus--nucleus collisions compared to that in pp collisions~\cite{Anastasia}.

\section{D-meson reconstruction}

The decays ${\rm D^0}\rightarrow {\rm K^-}\pi^{+}$, ${\rm D^+}\rightarrow {\rm K^-}\pi^{+}\pi^{+}$ and ${\rm D^{*+}}\rightarrow {\rm D^0}\pi^{+}$, and their charge conjugates, were reconstructed at mid-rapidity ($|y|<0.8$) in minimum-bias Pb--Pb collisions using the ALICE central barrel detectors. The D-meson selection was based on the precise reconstruction of the primary and secondary (decay) vertices, which is provided by the Inner Tracking System (ITS).
Charged pions and kaons were identified using the information provided by the Time Projection Chamber (TPC) and the Time Of Flight (TOF) detectors~\cite{Abelev:2014ffa}.
The reference proton--proton cross section at $\sqrt{s_{\rm NN}}=2.76$~TeV, needed to compute $R_{\rm AA}$, was obtained by a pQCD-based energy scaling of the $p_{\rm T}$-differential cross section measured at $\sqrt{s}=7$~TeV~\cite{ALICE:2011aa}. 

\section{D-meson nuclear modification factor}

\begin{figure}[!t]
\begin{minipage}[t]{0.48\textwidth}
\includegraphics[height=0.92\textwidth]{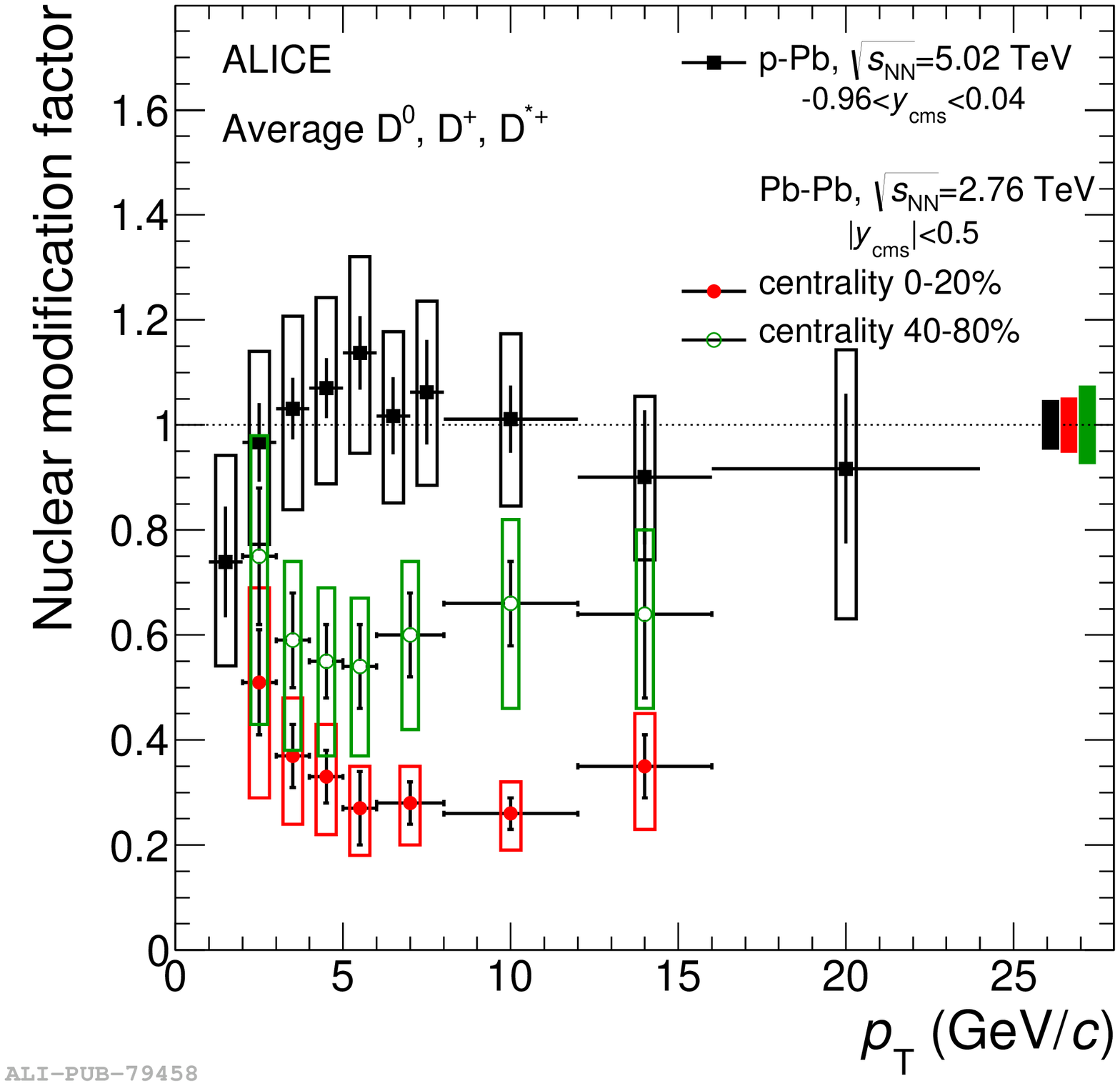}
\caption{\label{fig:DmesonRAARpA}Average $R_{\rm pPb}$ of prompt ${\rm D^0}$, ${\rm D^+}$ and ${\rm D^{*+}}$ mesons~\cite{Abelev:2014hha} compared to the prompt D-meson $R_{\rm AA}$ in Pb--Pb collisions in the 0--20\% and 40--80\% centrality classes~\cite{ALICE:2012ab}.}
\end{minipage}\hspace{1pc}%
\begin{minipage}[t]{0.48\textwidth}
\includegraphics[height=0.9\textwidth]{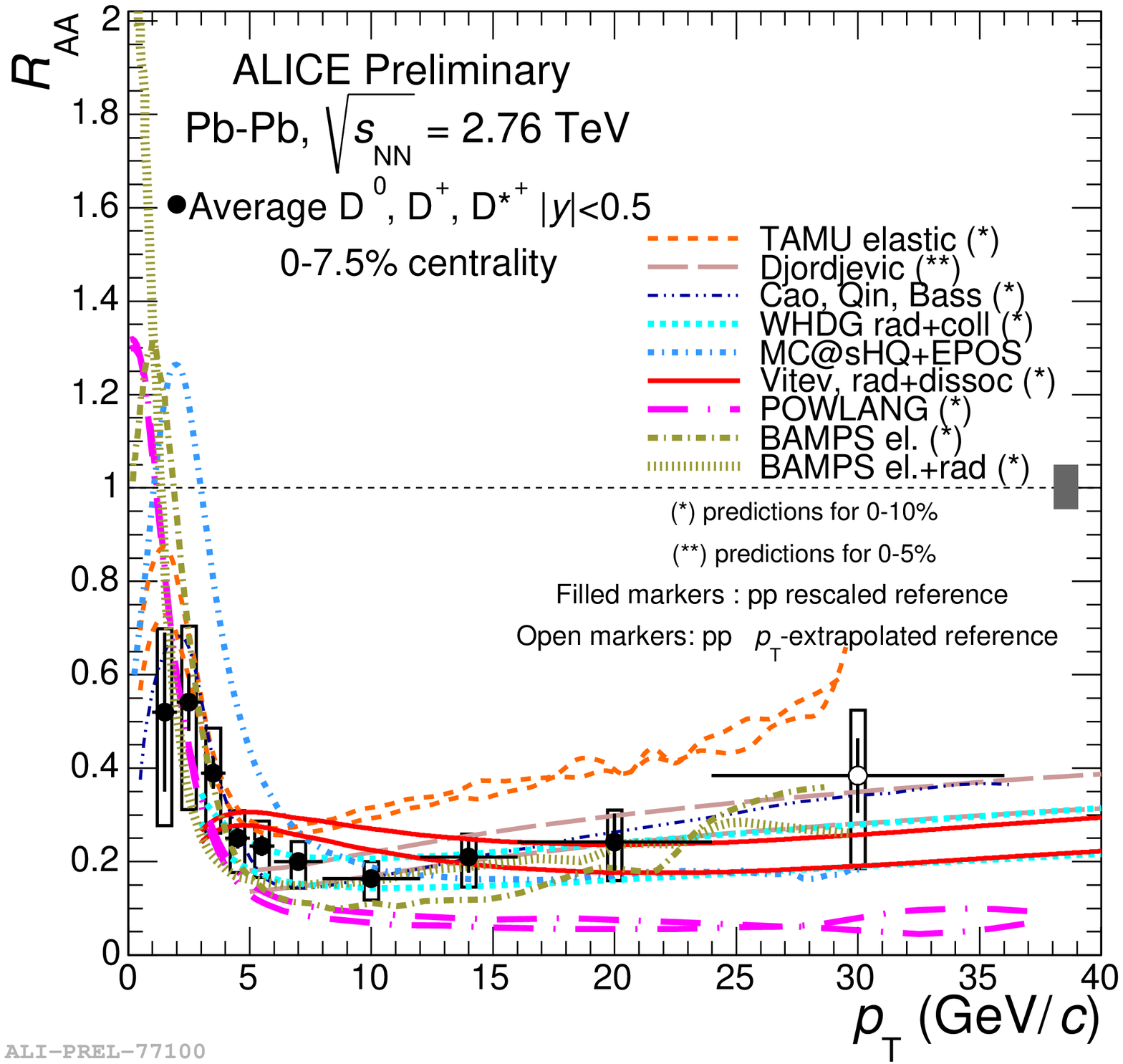}
\caption{\label{fig:DmesonRAACentral}Average prompt D-meson $R_{\rm AA}$ in Pb--Pb collisions in the 0--7.5\% centrality class compared to theoretical models including parton energy loss.}
\end{minipage} 
\end{figure}

A large suppression of the D-meson $R_{\rm AA}$ (factor 3-5) was observed for $p_{\rm T}>5$~GeV/$c$ in central Pb--Pb collisions at $\sqrt{s_{\rm NN}}=2.76~{\rm TeV}$ (Figure~\ref{fig:DmesonRAARpA})~\cite{ALICE:2012ab}. The comparison of $R_{\rm AA}$ with the D-meson nuclear modification factor measured in p--Pb collisions at $\sqrt{s_{\rm NN}}=5.02~{\rm TeV}$~\cite{Abelev:2014hha} shows that the expected cold nuclear matter effects are smaller than the uncertainties on $R_{\rm AA}$ for $p_{\rm T}>3~{\rm GeV/}c$. Therefore, the suppression observed in central Pb--Pb collisions is predominantly induced by final-state effects due to the presence of a hot and dense partonic medium. Figure~\ref{fig:DmesonRAACentral} shows the D-meson $R_{\rm AA}$ measured in Pb--Pb collisions in the centrality class 0--7.5\%, compared to theoretical models including charm interactions with the medium constituents.
The large suppression observed, e.g. of a factor 6 at $p_{\rm T}=10~{\rm GeV/}c$, is described by the models that include radiative and collisional heavy-quark energy loss.

\begin{figure}[!t]
\begin{minipage}[t]{0.48\textwidth}
\includegraphics[height=0.93\textwidth]{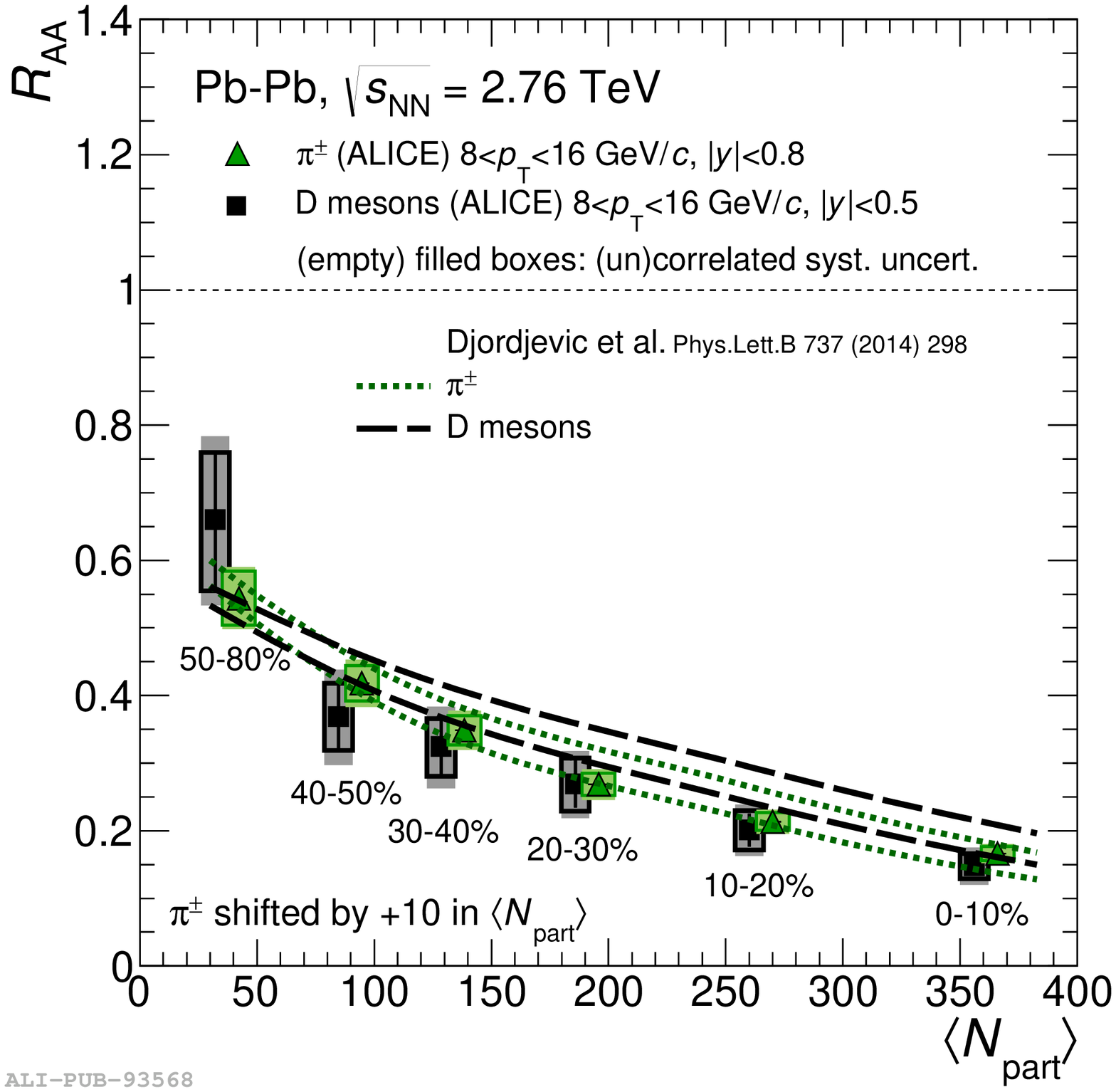}
\caption{\label{fig:DmesonPionRAA}$R_{\rm AA}$ of D mesons~\cite{Adam:2015nna} and charged pions~\cite{Abelev2014196} as a function of centrality compared to a pQCD model including mass dependent radiative and collisional energy loss~\cite{Djordjevic2014298}.}
\end{minipage}\hspace{1pc}%
\begin{minipage}[t]{0.48\textwidth}
\includegraphics[height=0.93\textwidth]{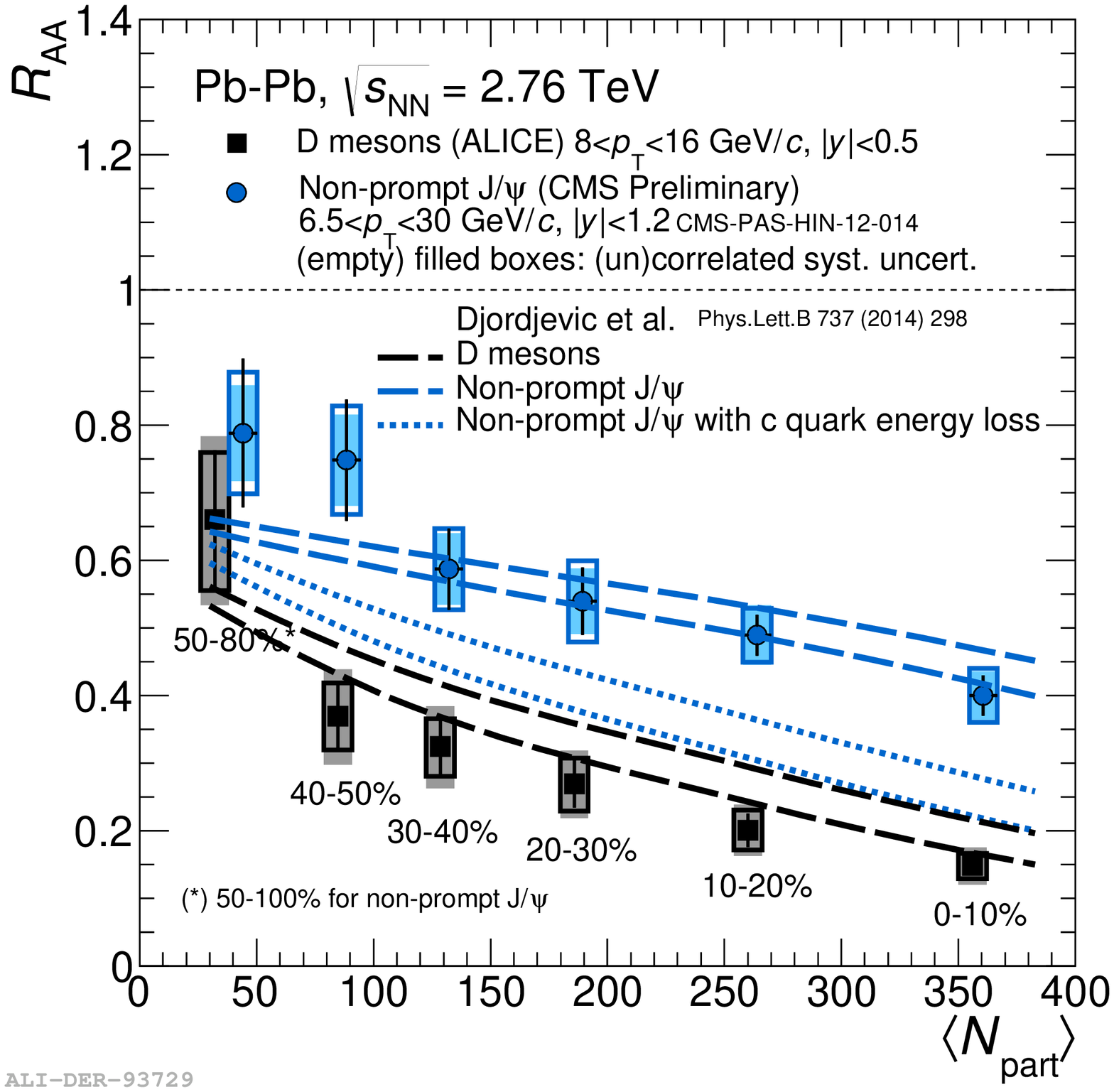}
\caption{\label{fig:DmesonJPsiRAA}D~\cite{Adam:2015nna} and non-prompt J/$\psi$ meson~\cite{CMSnonprompt} $R_{\rm AA}$ vs. centrality compared to a pQCD model including mass dependent radiative and collisional energy loss~\cite{Djordjevic2014298}.}
\end{minipage} 
\end{figure}

Figures~\ref{fig:DmesonPionRAA} and~\ref{fig:DmesonJPsiRAA} show the D-meson $R_{\rm AA}$ as a function of centrality (quantified in terms of the average number of participant nucleons in the Pb--Pb collision)~\cite{Adam:2015nna} along with the $R_{\rm AA}$ of charged pions~\cite{Abelev2014196} and non-prompt J/$\psi$ mesons measured by the CMS Collaboration~\cite{CMSnonprompt}, respectively. The focus here is on the study of the parton energy loss, thus, the results are presented for the high-$p_{\rm T}$ interval $8-16~{\rm GeV/}c$ for the pions and D mesons and for $6.5<p_{\rm T}<30~{\rm GeV/}c$ for J/$\psi$ from B-meson decays. The $R_{\rm AA}$ of charged pions and D mesons are compatible within uncertainties in all centrality classes. The average difference of the $R_{\rm AA}$ values of D mesons and non-prompt J/$\psi$ in the 0--10\% and 10--20\% centrality classes is larger than zero with a significance of 3.5~$\sigma$. A pQCD model including mass-dependent radiative and collisional energy loss~\cite{Djordjevic2014298} describes well the similarity of the D-meson and charged-pion $R_{\rm AA}$.
In this model, the colour-charge dependence of energy loss is compensated by the softer fragmentation and $p_{\rm T}$ spectrum of gluons with respect to those of c quarks leading to a very similar $R_{\rm AA}$ for D mesons and pions. This calculation results in a larger suppression of D mesons with respect to non-prompt J/$\psi$, in agreement with the data. The large difference in the two $R_{\rm AA}$ derives predominantly from the quark-mass dependence of parton energy loss, as demonstrated by the non-prompt J/$\psi$ $R_{\rm AA}$ calculated considering the c-quark mass in the energy loss of b quarks (dotted lines in Figure~\ref{fig:DmesonJPsiRAA})~\cite{Djordjevic2014298,PhysRevC.89.014905,He2013409}.

\section{D-meson azimuthal anisotropy}

\begin{figure}[!t]
\begin{minipage}[t]{0.49\textwidth}
\includegraphics[height=0.81\textwidth]{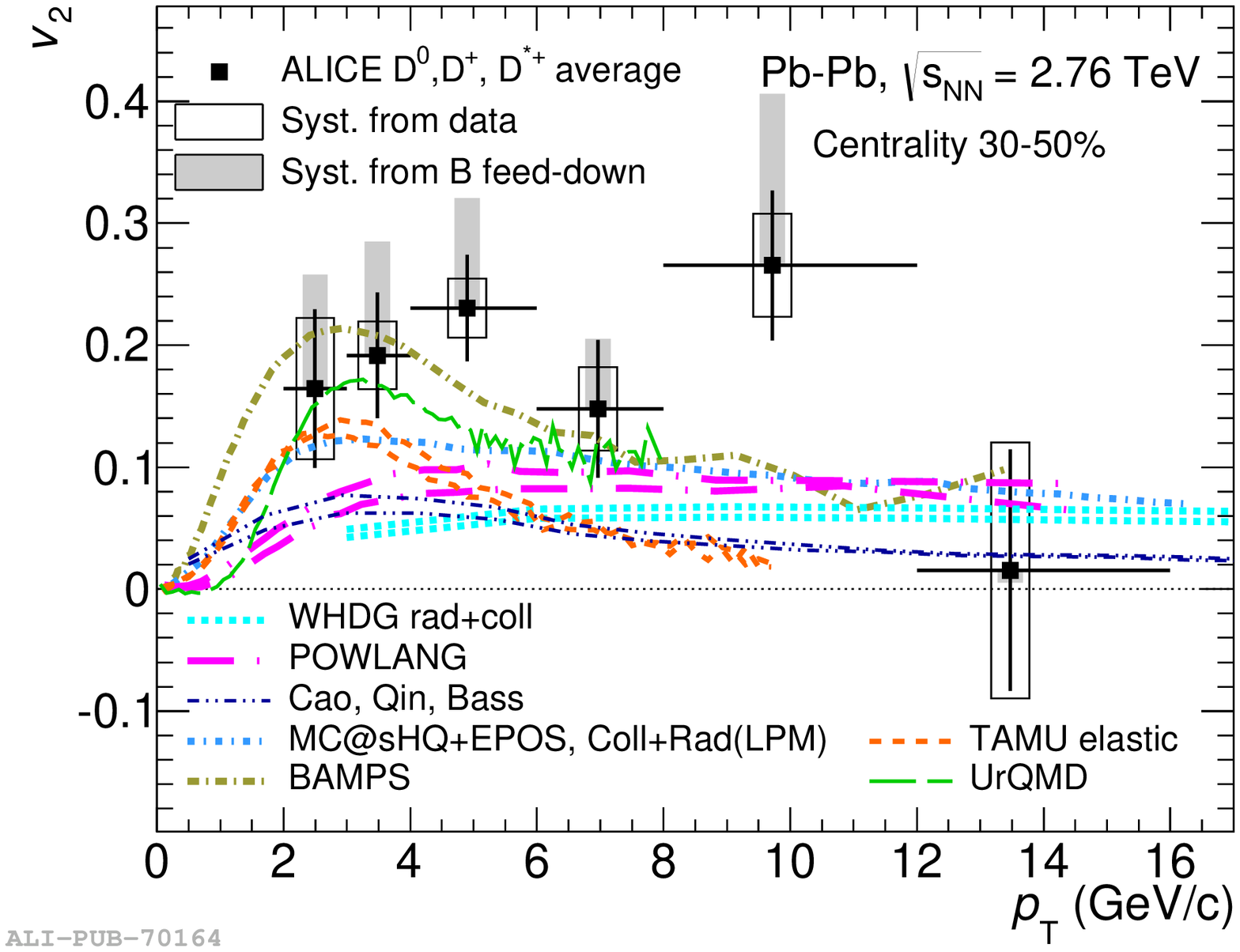}
\caption{\label{fig:Dmesonv2}Model comparison for the average D-meson $v_2$ in the 30--50\% centrality class~\cite{Abelev:2013lca,Abelev:2014ipa}.}
\end{minipage}\hspace{1pc}%
\begin{minipage}[t]{0.48\textwidth}
\centering
\includegraphics[height=0.85\textwidth]{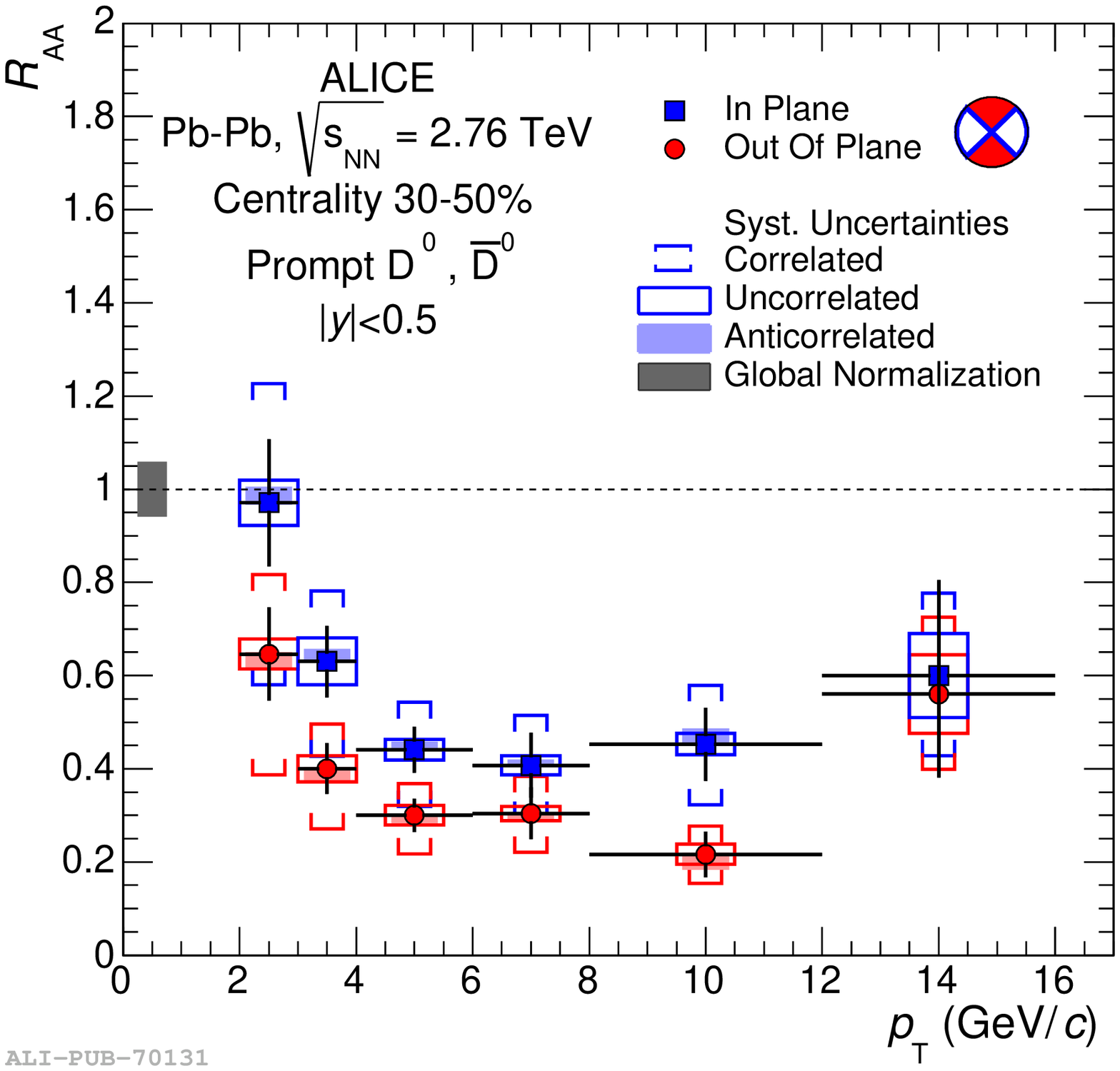}
\caption{\label{fig:DmesonRAAvsEP}${\rm D^0}$ $R_{\rm AA}$ in the 30--50\% centrality class in two $90^{\circ}$-wide orthogonal azimuthal intervals~\cite{Abelev:2014ipa}.}
\end{minipage} 
\end{figure}

$v_2$ is defined as the second coefficient of the Fourier expansion of the $p_{\rm T}$-dependent particle distribution ${\rm d}^2N/{\rm d}p_{\rm T}{\rm d}\varphi$ in azimuthal angle $\varphi$ relative to the reaction plane. It quantifies the momentum azimuthal anisotropy of final-state D mesons that might result from the spatial asymmetry of the initial parton distribution in non-central collisions, as a consequence of the collective expansion of the QGP. Figure~\ref{fig:Dmesonv2} shows the D-meson $v_2$ in Pb--Pb collisions in the 30--50\% centrality class~\cite{Abelev:2013lca,Abelev:2014ipa} compared to theoretical models including charm-quark interactions with the medium constituents.
The value of $v_2$ is larger than zero in the interval $2<p_{\rm T}<6$~GeV/$c$ with a significance of about 5~$\sigma$ and it is best described by models including mechanisms that transfer to the charm quarks the elliptic flow of the medium during the system expansion (namely collisional processes and hadronisation by recombination with light quarks). A smaller suppression is observed in the in-plane direction (Figure~\ref{fig:DmesonRAAvsEP})~\cite{Abelev:2014ipa}; the ordering $R_{\rm AA}^{\rm out-of-plane}<R_{\rm AA}^{\rm in-plane}$ is equivalent to the observation of $v_2>0$ and, thus, consistent with the expectations from collective flow.

\section{Conclusions}
The results obtained by ALICE using the data from the LHC Run-1 (2010--2013) indicate a strong suppression of the D-meson production in central Pb--Pb collisions for $p_{\rm T}>3~{\rm GeV/}c$, which is mainly due the interactions of heavy quarks with the hot and dense medium. 
The smaller $R_{\rm AA}$ observed for D mesons with respect to non-prompt J/$\psi$ confirms the mass-dependent nature of the energy-loss mechanisms.
The non-zero $v_2$ of D mesons and the azimuthal dependence of the ${\rm D^0}$ $R_{\rm AA}$ indicate that, during the collective expansion of the medium, the interactions between its constituents and the charm quarks transfer to the latter information on the azimuthal anisotropy of the system. During the LHC Run-2 we expect to collect a data sample larger by a factor 5-10 with respect to Run-1, depending on collision centrality. It will be, thus, possible to measure the D-meson $R_{\rm AA}$ and $v_2$ with a better precision and in an extended $p_{\rm T}$ range.

\section*{References}
\bibliography{iopart-num}

\providecommand{\newblock}{}
\begin{thebibliography}{10}
\expandafter\ifx\csname url\endcsname\relax
  \def\url#1{{\tt #1}}\fi
\expandafter\ifx\csname urlprefix\endcsname\relax\def\urlprefix{URL }\fi
\providecommand{\eprint}[2][]{\url{#2}}

\bibitem{Baier1997265}
Baier R, Dokshitzer Y, Mueller A, Peign\'e S and Schiff D 1997 {\em Nuclear
  Physics B\/} {\bf 484} 265--282 ISSN 0550-3213

\bibitem{PhysRevD.44.1298}
Braaten E and Thoma M~H 1991 {\em Phys. Rev. D\/} {\bf 44}(4) 1298--1310

\bibitem{PhysRevD.69.114003}
Armesto N, Salgado C~A and Wiedemann U~A 2004 {\em Phys. Rev. D\/} {\bf 69}(11)
  114003

\bibitem{Batsouli200326}
Batsouli S, Kelly S, Gyulassy M and Nagle J 2003 {\em Physics Letters B\/} {\bf
  557} 26--32 ISSN 0370-2693

\bibitem{Greco2004202}
Greco V, Ko C and Rapp R 2004 {\em Physics Letters B\/} {\bf 595} 202--208 ISSN
  0370-2693

\bibitem{Andronic200336}
Andronic A, Braun-Munzinger P, Redlich K and Stachel J 2003 {\em Physics
  Letters B\/} {\bf 571} 36--44 ISSN 0370-2693

\bibitem{Anastasia}
Barbano A (for the ALICE Collaboration) 2015 {\em These proceedings\/}

\bibitem{Abelev:2014ffa}
Abelev B~B {\em et~al.\/} (ALICE) 2014 {\em Int. J. Mod. Phys.\/} {\bf A29}
  1430044 (\textit{Preprint} \eprint{1402.4476})

\bibitem{ALICE:2011aa}
Abelev B~B {\em et~al.\/} (ALICE) 2012 {\em JHEP\/} {\bf 01} 128
  (\textit{Preprint} \eprint{1111.1553})

\bibitem{Abelev:2014hha}
Abelev B~B {\em et~al.\/} (ALICE) 2014 {\em Phys. Rev. Lett.\/} {\bf 113}
  232301 (\textit{Preprint} \eprint{1405.3452})

\bibitem{ALICE:2012ab}
Abelev B~B {\em et~al.\/} (ALICE) 2012 {\em JHEP\/} {\bf 09} 112
  (\textit{Preprint} \eprint{1203.2160})

\bibitem{Adam:2015nna}
Adam J {\em et~al.\/} (ALICE) 2015  (\textit{Preprint} \eprint{1506.06604})

\bibitem{Abelev2014196}
Abelev B~B {\em et~al.\/} (ALICE) 2014 {\em Physics Letters B\/} {\bf 736}
  196--207 ISSN 0370-2693

\bibitem{Djordjevic2014298}
Djordjevic M, Djordjevic M and Blagojevic B 2014 {\em Physics Letters B\/} {\bf
  737} 298--302 ISSN 0370-2693

\bibitem{CMSnonprompt}
{CMS Collaboration} (CMS) 2012 {\em CMS-PAS-HIN-12-014\/}

\bibitem{PhysRevC.89.014905}
Nahrgang M, Aichelin J, Gossiaux P~B and Werner K 2014 {\em Phys. Rev. C\/}
  {\bf 89}(1) 014905

\bibitem{He2013409}
He M, Fries R~J and Rapp R 2013 {\em Nuclear Physics A\/} {\bf 910–911}
  409--412 ISSN 0375-9474

\bibitem{Abelev:2013lca}
Abelev B~B {\em et~al.\/} (ALICE) 2013 {\em Phys.Rev.Lett.\/} {\bf 111} 102301
  (\textit{Preprint} \eprint{1305.2707})

\bibitem{Abelev:2014ipa}
Abelev B~B {\em et~al.\/} (ALICE) 2014 {\em Phys. Rev.\/} {\bf C90} 034904
  (\textit{Preprint} \eprint{1405.2001})

\end{thebibliography}

\end{document}